\documentclass[12pt]{iopart}

\usepackage{epsfig}

\begin{document}

\title{Systematic Comparison of Jet Energy-Loss Schemes
in a 3D hydrodynamic medium}
\author{S.A.~Bass$^1$, C.~Gale$^2$, A. Majumder$^1$, C.~Nonaka$^3$, G.-Y.~Qin$^2$,  
T.~Renk$^{4,5}$ and J. Ruppert$^{2,6}$ }
\address{$^1$ Department of Physics, Duke University, 
             Durham, North Carolina 27708-0305, USA}
\address{$^2$Department of Physics, McGill University, H3A 2T8, 
		Montreal, Quebec, Canada}
\address{$^3$Department of Physics, Nagoya University, Nagoya 464-8602, Japan}  		
\address{$^4$Department of Physics, PO Box 35 FIN-40014 University of 
		Jyv\"{a}skyl\"{a}, Finland}
\address{$^5$Helsinki Institute of Physics, PO Box 64 FIN-00014 University 
		of Helsinki, Finland}
\address{$^6$Institut f\"ur Theoretische Physik,, J.W. Goethe University Frankfurt, 
         Max-von-Laue-Str. 1, D-60438 Frankfurt am Main, Germany}          

\ead{bass@phy.duke.edu}

\submitto{\JPG}

\begin{abstract}
We utilize a 3D hydrodynamic model to provide the medium evolution 
for a systematic comparison of jet energy-loss calculations in the 
BDMPS/ASW, HT and AMY approaches. 
We find that the parameters of all three calculations
can be adjusted to provide a good description of inclusive data on $R_{AA}$ versus 
transverse momentum. However, we do observe slight differences in their 
predictions for the
azimuthal angular dependence of $R_{AA}$ vs. $p_T$. We also note that the value
of the transport coefficient $\hat{q}$ needed in the three approaches to describe
the data differs significantly.
\end{abstract}


Experiments at the 
Relativistic Heavy Ion Collider (RHIC) 
 have established a significant suppression
of high-$p_T$ hadrons produced in central A+A collisions compared to those
produced in peripheral A+A or binary scaled p+p reactions, indicating
a strong nuclear medium effect \cite{Adcox:2001jp,Adler:2002xw}, commonly referred to as {\em jet-quenching}.
Within the framework of perturbative QCD, the leading process of energy 
loss of a fast parton is gluon radiation induced by multiple soft collisions 
of the leading parton or the radiated gluon with color charges in the
quasi-thermal medium \cite{Gyulassy:1993hr,Baier:1996kr,Zakharov:1997uu}.

Over the past two years, a large amount of 
jet-quenching related experimental data has become available, 
including but not limited to
the nuclear modification factor $R_{AA}$, the elliptic
flow $v_2$ at high $p_T$ (as a measure of the
azimuthal anisotropy of the jet cross section) and
a whole array of high $p_T$ hadron-hadron correlations.
Computations of such jet modifications have acquired a
certain level of sophistication regarding the
incorporation of the partonic processes involved.
However, most of these calculations have been utilizing
simplified models for the underlying soft medium,
e.g. assuming a simple density distribution and its
variation with time. Even in more elaborate setups, most jet
quenching calculations assume merely a one- or two-dimensional Bjorken
expansion.

The availability of a three-dimensional hydrodynamic evolution
code \cite{Nonaka:2006yn}  allow
for a much more detailed study of jet interactions in a
longitudinally and transversely expanding medium. 
The variation of the gluon density as a function of space and time in these approaches
is very different from  that in a simple Bjorken expansion.
A previous calculation in this direction \cite{Hirano:2002sc,Hirano:2003pw}
estimated the effects of 3-D expansion on the $R_{AA}$.
However, this approach treated the
energy loss of jets in a rather simplified manner, with the energy loss $dE/dx$ exhibiting
a simple linear dependence on the product of hard scattering cross section and
gluon density (as a function of temperature extracted from the hydrodynamic simulation).
Over the past year we have  utilized our 
3-D hydrodynamic model to provide the time-evolution of the medium produced at RHIC 
for jet energy-loss calculations performed in the BDMPS/ASW \cite{Renk:2006sx}, 
Higher Twist \cite{Majumder:2007ae} and AMY \cite{Qin:2007zz} approaches.  
In each of the three efforts, the inclusive as well as the azimuthally differential
nuclear suppression factor $R_{AA}$ of pions was studied as a function of their 
transverse momentum $p_T$. In addition, the influence of collective flow, variations 
in rapidity, and energy-loss in the hadronic phase were addressed for the selected 
approaches. For details regarding the implementation of the energy-loss schemes
and their interface to the hydrodynamic medium, we refer the reader to the publications
cited above. Here we shall focus on a systematic comparison between the three
approaches, utilizing the same hydrodynamic medium evolution as well as the same
structure and fragmentation functions for calculating the initial state and final
high-$p_T$ hadron distributions.

\begin{figure}[tb]
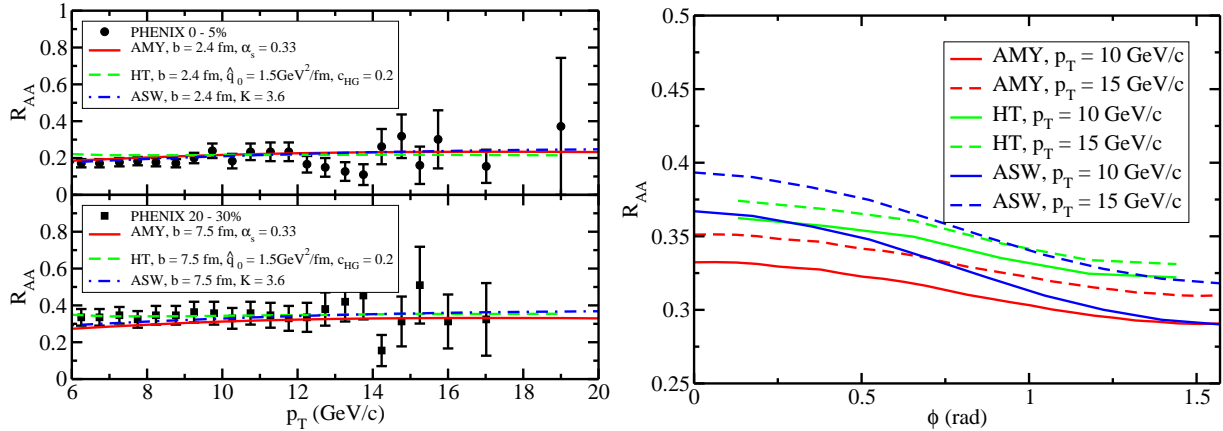
   
\centerline{\epsfig{file=RAA_centrality.eps,width=8cm}
\epsfig{file=RAA_fixed_pT2.eps,width=8cm}}
\caption{Left: Nuclear modification factor $R_{AA}$ in $Au$-$Au$ collisions at 0-5\% (top) 
and 20-30\% (bottom) centrality calculated in the ASW, HT and AMY approaches compared to data from PHENIX~\cite{Shimomura:2005en}. Right: $R_{AA}$ as a function of azimuthal angle at $p_T=10$~GeV/c
(solid line) and $p_T=15$~GeV/c (dashed line) for all three approaches in the 20-30\% centrality bin.}
\label{fig1}
\end{figure}

The left frame of figure~\ref{fig1} shows the nuclear modification factor $R_{AA}$ in $Au$-$Au$
collisions at 0-5\% (top)  and 20-30\% (bottom) centrality calculated in the ASW, HT and AMY approaches compared to data from PHENIX~\cite{Shimomura:2005en}. As can be seen, the parameters
for all three approaches (initial maximal value for the transport coefficient $\hat{q}_0$
or coupling
constant $\alpha_s$ in the AMY case) can be adjusted such that the approaches are able to 
describe the centrality dependence of the nuclear modification factor reasonably well. 
For a gluon jet, the
values are $\hat{q}_0 \approx 3.4$~GeV$^2$/fm for the HT approach,
$\hat{q}_0 \approx 20$~GeV$^2$/fm for the ASW formalism and $\alpha_S \approx 0.33$ for
the AMY approach, which can be converted into a value of $\hat{q}_0 \approx 5.5$~GeV$^2$/fm.
Note that  the ASW value for $\hat{q}_0$ at $\tau=0.6$~fm/c and $\epsilon_0=55$~GeV/fm$^3$
lies a factor of 3.6 higher
than the Baier estimate for an ideal QGP, $\hat{q} \approx 2 \cdot \epsilon^{3/4}$
\cite{Baier:2002tc},  while the AMY value is in line with the ideal QGP estimate
and the HT calculation lies about a factor of 1.6 below that estimate.
The large difference in $\hat{q}_0$ values between HT and ASW has been pointed out previously.
However, we find that a factor of two can be accounted for by the use of different scaling
prescriptions (temperature vs. energy-density) with which the medium is coupled to
the transport coefficient -- this will be discussed in greater detail in a forthcoming
publication.

\begin{figure}[tb]
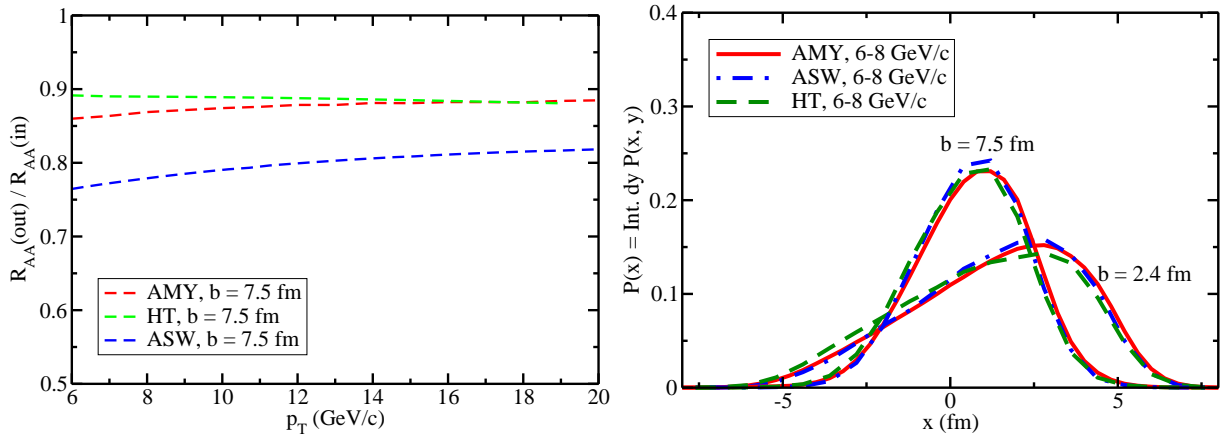
   
\centerline{\epsfig{file=RAA_out_in_ratio.eps,width=8cm}
\epsfig{file=RAA_Px.eps,width=8cm}}
\caption{Left: Ratio $R_{AA}$ for out of plane vs. in plane emission as a function of $p_T$ 
at b=7.5 fm impact parameter for all three approaches. Right: escape probability of a hadron
with 6-8 GeV/c transverse momentum moving along the positive $x$-axis in the transverse
plane as a function of $x$.}
\label{fig2}
\end{figure}

We find that slight variations appear between the approaches when $R_{AA}$ is 
studied as a function of azimuthal angle. This can be seen in the right frame of figure~\ref{fig1}
where $R_{AA}$ is plotted as a function of azimuthal angle at $p_T=10$~GeV/c
(solid line) and $p_T=15$~GeV/c (dashed line) for all three approaches in the 20-30\% centrality bin.
In order to quantify the difference between the three approaches we calculate the 
ratio of the out of plane $R_{AA}$ over the in plane $R_{AA}$ as a function of transverse
momentum -- this is shown in the left frame of figure~\ref{fig2}. We find that AMY and HT
exhibit the same peak to valley ratio, even though the absolute values for $R_{AA}$ differ
by approximately 10\%. The ASW calculation systematically shows a stronger azimuthal dependence
than the HT and AMY calculations - the cause of which will require a more detailed analysis
to determine.

In order to investigate the spatial response of the jet energy-loss schemes to the medium
the right frame of figure~\ref{fig2} shows the escape probability of a hadron with
a transverse momentum between 6 and 8 GeV/c originating
from a quenched jet moving in the positive $x$ direction in the transverse plane as a 
function of of its production vertex along the $x$-axis. Mathematically this quantity is defined as:
\begin{equation}
P(x)= \int {\rm dy} \, T_{AB}(x,y) \cdot R_{AA}(x,y)\,/{\int {\rm dx dy} \, T_{AB}(x,y) \cdot R_{AA}(x,y)}
\end{equation}
It is remarkable how well the three different approaches agree with each other in this
quantity. Since the same hard scattering probability was used as input in all three
cases, the agreement
in $P(x)$ really shows that all three approaches yield the same suppression factor 
as a function of production vertex of the hard probe, i.e. that they probe the 
density of the medium in the same way.

In summary, our comparison shows
that under identical conditions (i.e. same medium evolution, same choice
of parton distribution functions, scale etc.) all three jet energy-loss schemes yield 
very similar results. This finding is very encouraging since it indicates that the technical
aspects of the formalisms are well under control. However, we need to point out that there
still exist significant differences regarding the extracted value for the transport coefficient 
$\hat{q}_0$, which have yet to be fully understood.
 
\ack   
This work was supported in part by an Outstanding Junior Investigator
Award  from the U.~S.~Department of Energy (grant DE-FG02-05ER41367). TR was supported in part
by the Academy of Finland, Project 206024 and CG, GQ and JR acknowledge support by the Natural Sciences and Engineering Research Council of Canada.

\section*{References}

\bibliographystyle{iopart-num}

\bibliography{/Users/bass/Publications/SABrefs}

\end{document}